\documentclass[twocolumn,showpacs,preprintnumbers,amsmath,amssymb,dvipdfmx]{revtex4-1}

\usepackage{graphicx}
\usepackage{dcolumn}
\usepackage{bm}

\begin{document}

\title{
Vison-Majorana complex zero-energy resonance in Kitaev's spin liquid
}

\author{Masafumi Udagawa}%
\affiliation{%
Department of Physics, Gakushuin University, Mejiro, Toshima-ku, Tokyo 171-8588, Japan
}%

\date{\today}

\begin{abstract}
 We study the effect of site dilution in Kitaev's model.
We derive an analytical solution of the dynamical spin correlation functions for arbitrary configurations of $Z_2$ fluxes.
By incorporating this solution into classical Monte Carlo scheme, we address how a site vacancy affects the experimental observables,
such as the static spin susceptibility and the spin lattice relaxation rate, $1/T_1$.
As a result, we found an enhancement of dynamical magnetic response in the vicinity of vacancy, which leads to Friedel-like oscillation in local $1/T_1$,
in contrast to limited influences on the static susceptibility. Furthermore, we found a sharp zero-energy peak in the magnetic excitation spectrum,
which is attributed to the Vison \& Majorana zero mode trapped near the site vacancy. This zero mode can be interpreted as fractionalized spin hole into an Ising triplet with differentiated magnetic axes, which leads to the characteristic temperature and field-orientational dependence of $1/T_1$.
\end{abstract}

\maketitle
Kitaev's honeycomb model is drawing considerable attention as a promising stage to realize quantum spin liquid (QSL) state~\cite{kitaev2006anyons}.
Theoretical proposals have been made for its realization in actual materials~\cite{jackeli2009mott,rau2014generic,winter2017models,plumb2014alpha,singh2010antiferromagnetic,trebst2017kitaev,PhysRevLett.108.127204,PhysRevLett.108.127203},
and several compounds are, indeed, intensively studied as promising candidates~\cite{kitagawa2018spin,banerjee2016proximate,banerjee2017neutron,2017NatPh..13.1079D}. 

While the realization of Kitaev's QSL is strongly desired, its experimental identification is not straightforward.
Absence of magnetic ordering down to lowest measurable temperatures is a promising signal, however, 
ultimately it is impossible to prove the absence of spontaneous symmetry breaking at real zero temperature. 
Given that the QSL ground state is featureless in terms of local observables, excitations, and especially their fractionalized nature may
provide better diagnose into QSL realization. In Kitaev's spin liquid, the original spin degrees of freedom is fractionalized, and the low-energy physics is described by 
two species of elementary excitations: itinerant c-Majoranas and Visons~\cite{kitaev2006anyons}.

Indeed, fractionalization results in several experimental consequences.
Most typically, dynamical response shows continuous rather than sharp resonant spectra, reflecting that a change of unit quantum number
gives rise to multi-particle process as a result of fractionalization~\cite{knolle2014dynamics,PhysRevB.92.115127}.
In fact, regarding Kitaev candidate material, $\alpha$-RuCl$_3$, continuous spectra were observed by inelastic neutron scattering~\cite{banerjee2016proximate,banerjee2017neutron,2017NatPh..13.1079D}, and it motivates intensive studies on the dynamics of this system~\cite{yoshitake2016fractional,yamaji2016clues,PhysRevB.96.024438,yoshitake2017temperature,samarakoon2018classical,ducatman2018magnetic,suzuki2018effective}.

However, there are still several objections regarding the interpretation of continuum as evidence for fractionalization.
On one hand, continuum may be a natural consequence of fractionalization, but on the other hand, it does not necessarily require fractionalization. For example, multi-magnon process may be considered as another possible mechanism, and indeed, such possibility was already discussed~\cite{winter2017breakdown}. 
Considering this ambiguity, it is desirable, if clearer evidence for fractionalization, based on resonance rather than continuum, is available.

Another important issue is the detection of Vison. Vison could be a main player in topological quantum computation~\cite{nayak2008non}, one of the most attractive applications of QSL.
Nevertheless, so far, there are no experimental tests to identify Vison.
It is desirable, if there is a physical phenomenon available that indicates the existence of Vison.

These unresolved issues motivate us to study the effect of impurity in Kitaev's spin liquid.
Impurity usually conceals rather than clarifies the true character of the system. However, in the context of spin liquid, 
or the systems of topological nature, impurity 
has traditionally served as useful tools to characterize their ground states.
So far, several types of impurities have been studied in Kitaev's model~\cite{willans2010disorder,willans2011site,PhysRevLett.105.117201,petrova2013unpaired,PhysRevB.90.134404,andrade2014magnetism,vojta2016kondo,PhysRevB.92.014403,PhysRevB.85.054204}, and the effect of atomic substitutions have been examined experimentally~\cite{PhysRevB.89.241102,PhysRevB.91.180406,PhysRevB.89.245113,PhysRevB.88.220414,PhysRevB.92.134412,PhysRevLett.119.237203}.
In this work, we will address the impurity effect in all range of temperatures, and try to provide a useful guide to experimental studies.

In this work, we adopt the Kitaev's model defined on a honeycomb lattice:
\begin{align}
\mathcal{H} = -\sum\limits_{i\in\Omega}J_xS_i^xS_{i_x}^x + J_yS_i^yS_{i_y}^y + J_zS_i^zS_{i_z}^z.
\end{align}
Here, $S_i^{\alpha}$ is the $\alpha$-th component of spin-$1/2$ operator, defined at site $i$. It interacts with its three neighboring spins at $i_{\alpha} (\alpha=x, y, z)$ through the Ising interactions with easy axis, $\alpha$. The summation of $i$ is taken over the sites on one of the sublattices, $\Omega$.
We consider this Hamiltonian in a system of $N_{\rm site}=2\times N\times N$ sites, 
assuming periodic boundary condition, as shown in Fig.~\ref{Fig.1} (a).
In this work, we focus on the case of ferromagnetic isotropic coupling, $J_x=J_y=J_z=1$. We also set $\hbar=k_{\rm B}=1$.

\begin{figure}[h]
\begin{center}
\includegraphics[width=0.49\textwidth]{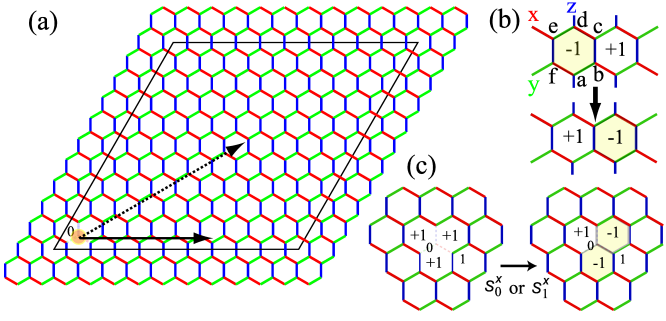}
\end{center}
\caption{\label{Fig.1} 
(color online). (a) Lattice geometry of the system under consideration. We consider the sites within a thin closed line with periodic boundary condition identifying the two pairs of opposite sides. The total number of sites is set $N_{\rm site}=2\times N\times N$ with $N=10$. We introduce a site vacancy at site $0$. 
 Two directions of high symmetry are shown with solid and dashed arrows, along which the local observables will be plotted in Fig.~\ref{Fig.2}.
(b) Schematic picture of $Z_2$ flux, defined as $W_p=\sigma_a^z\sigma_b^x\sigma_c^y\sigma_d^z\sigma_e^x\sigma_f^y$. Operation of $S_i^{\alpha}$ reverses a pair of $Z_2$ fluxes on the both sides of $\alpha$-bond, extending from site $i$. The case with $i=a$ and $\alpha=z$ is shown here. 
(c) The schematic figure of zero-energy Vison-pair creation in the vacancy hole, by the application of $S_0^x$ or $S_1^x$.
}
\end{figure}

By taking Majorana expression $S_i^{\alpha}=\frac{i}{2}c_ib_i^{\alpha}$, the Hamiltonian reads
\begin{align}
\mathcal{H}[\{W_p\}] = \frac{i}{4}\sum\limits_{i\in\Omega}\sum_{\alpha}u_i^{\alpha}c_ic_{i_{\alpha}}\equiv\frac{i}{4}c_kA_{kk'}c_{k'},
\label{cHamiltonian}
\end{align}
where a set of gauge fields, $u_i^{\alpha}=ib_i^{\alpha}b_{i_{\alpha}}^{\alpha}$, are introduced. We fix one representative set of $\{u_i^{\alpha}\}$ for each configurations of gauge-invariant $Z_2$ fluxes, $\{W_p\}$.
To accomplish the projection onto physical space, in addition to this gauge fixing, we need to choose physical fermion parity. 
Namely, by defining complex fermion $f_i\equiv\frac{1}{2}(ic_i+c_{i_z})$ for $i\in\Omega$, we define the fermion parity (f-parity): $(-1)^F\equiv(-1)^{{\sum_{i\in\Omega}}f_i^{\dag}f_i}$, which commutes with the Hamiltonian, eq.~(\ref{cHamiltonian}). With the lattice geometry defined above [Fig.~\ref{Fig.1} (a)], only the states with specific f-parity,
\begin{align}
(-1)^{F}=(-1)^{F_{\rm ph}}\equiv\prod_{i\in\Omega}\prod_{\alpha=x,y,z}u_i^{\alpha},
\label{fparity}
\end{align}
turns out to be physical, according to the criterion given in ref.~\cite{pedrocchi2011physical}.
For a simple derivation of this formula, see Supplemental material.

In the ground state, the $Z_2$ fluxes take $W_p=+1$, uniformly.
However, at finite temperatures, Visons, namely the hexagons carrying $W_p=-1$ are thermally excited [Fig.~\ref{Fig.1} (b)].
In order to study the finite-temperature property, we use the classical Monte Carlo method by sampling $\{W_p\}$. 
In each realization of $Z_2$ fluxes, we choose the representative set of $u_i^{\alpha}$, and determine the gauge-fixed form of quadratic c-Majorana Hamiltonian through eq.~(\ref{cHamiltonian}).
We then diagonalize the matrix $iA$ and obtain a set of eigenvalues, $\{\varepsilon^{(m)}\}$, and corresponding eigenvector, $\{v_j^{(m)}\}$.
The expectation values of physical observables are given by
\begin{align}
\langle\mathcal{O}\rangle &= \sum_{\{W_p\}}\frac{Z\{W_p\}}{Z}\frac{{\rm Tr}P_F\mathcal{O}e^{-\beta\mathcal{H}[\{W_p\}]}}{Z\{W_p\}},
\label{Observables}
\end{align}
with
$Z\{W_p\}\equiv{\rm Tr}P_Fe^{-\beta\mathcal{H}[\{W_p\}]}, Z=\sum_{\{W_p\}}Z\{W_p\}$.
Here, to impose correct f-parity in the Monte-Carlo process, we have introduced the f-parity projection operator,
\begin{align}
P_F\equiv\frac{1}{2}[1+(-1)^{F_{\rm ph}}e^{i\pi\sum_{i\in\Omega}f_i^{\dag}f_i}],
\end{align}
where $(-1)^{F_{\rm ph}}$ is given by eq.~(\ref{fparity}). 

In this work, we focus on the correlation function, 
\begin{eqnarray}
\int_0^{\infty}dt\ \langle S_j^{\alpha}(t)S_{j'}^{\alpha'}\rangle e^{i(\omega+i\delta)t}=\delta_{\alpha\alpha'}\Psi^{\alpha}_{jj'}(\omega).
\label{correlation_omega}
\end{eqnarray}
Here, the real and imaginary part of this correlation function can be associated with observable quantities. The local magnetic susceptibility, $\chi_i$; the moment induced at site $i$ divided by the applied uniform magnetic field, is given by
\begin{align}
\chi_i/(g\mu_{\rm B})^2=-\frac{2}{3}{\rm Im}\sum_{\alpha=x,y,z}(\Psi^{\alpha}_{ii}(0) + \Psi^{\alpha}_{ii_{\alpha}}(0)),
\end{align}
where we assumed the field parallel to $[111]$. 

The real part is associated with the dynamical magnetic susceptibility, $\chi_{jj'}^{\alpha\alpha}(\omega)$ as
\begin{eqnarray}
\lim_{\omega\to0}{\rm Re}\Psi^{\alpha}_{jj'}(\omega)=\frac{T}{2\pi}\lim_{\omega\to0}\frac{\chi_{jj'}^{\alpha\alpha}(\omega)}{\omega}\equiv\bigl(\frac{1}{T_1}\bigr)_j^{\alpha}.
\label{Realpart_correlation}
\end{eqnarray}
Usually, the spin-lattice relaxation rate, $1/T_1$, at some nucleus, is associated with the summation of the righthand side of eq.~(\ref{Realpart_correlation}) over the neighboring atoms. In this letter, we regard $\bigl(\frac{1}{T_1}\bigr)_j^{\alpha}$ defined above as local $1/T_1$, representing its value near the site $j$. Both $\chi_i$ and $\bigl(\frac{1}{T_1}\bigr)_j^{\alpha}$ can be observable through the NMR, which has been widely conducted for a number of Kitaev candidate materials~\cite{kitagawa2018spin,janvsa2018observation,zheng2017j,baek2017evidence}.

We obtained an analytical formula of the correlation function for each realization of $Z_2$ fluxes. Performing an average over the configuration of $\{W_p\}$, we can write
\begin{widetext}
\begin{align}
&\langle S_j^{\alpha}(t)S_{j'}^{\alpha}(0)\rangle = \frac{1}{2\sum_{\{W_p\}} Z[\{W_p\}]}\sum_{\{W_p\}}\Bigl(\sqrt{{\rm det}(1 + e^{-(\beta-it)\cdot iA}e^{-it\cdot iA^{(j)}})} \Bigl[\frac{1}{1 + e^{-(\beta-it)\cdot iA}e^{-it\cdot iA^{(j)}}}e^{-(\beta-it)\cdot iA}\Bigr]_{j'j}\nonumber\\
&\hspace{0.2cm}-(-1)^{F_{\rm ph}}\sqrt{{\rm det}(1 - e^{-(\beta-it)\cdot iA}e^{-it\cdot iA^{(j)}})} \Bigl[\frac{1}{1 - e^{-(\beta-it)\cdot iA}e^{-it\cdot iA^{(j)}}}e^{-(\beta-it)\cdot iA}\Bigr]_{j'j}\Bigr)(\delta_{j'j} - iu_j^{\alpha}\delta_{j'j_{\alpha}})
\label{analyticalformula}
\end{align}
\end{widetext}
Here, $A$ is the $N_{\rm site}\times N_{\rm site}$ matrix defined in eq.~(\ref{cHamiltonian}), whose elements depend on the fixed gauge fields, $\{u_i^{\alpha}\}$, corresponding to the $Z_2$ flux configuration, $\{W_p\}$. Meanwhile, $A^{(j)}$ is the matrix with modified gauge fields, after the operation of $S^{\alpha}_j$, which changes the $Z_2$ flux configuration to $\{W'_p\}$, by reversing a pair of $W_p$'s on the both sides of the $\alpha$-bond, extending from the site $j$ [Fig.~\ref{Fig.1}(b)].
The formula (\ref{analyticalformula}) is the first result of this work. 
The advantage of the expression (\ref{analyticalformula}) is that it allows us to access the real-time correlation functions directly.
From this expression, we first calculate $\langle S_j^{\alpha}(t)S_{j'}^{\alpha}(0)\rangle$ at each discretized times until $t=200$ as a upper cut of the time integral in eq.~(\ref{correlation_omega}), then
by making Fourier transformation to obtain the dynamical correlation function, $\Psi^{\alpha}_{jj'}(\omega)$. 
We perform the calculation for the system with $N=10$, i.e. 200 sites. Typically, we use 40000 Monte Carlo steps for the lowest temperatures.
For further details of the calculation, see Supplemental material.

Now, let us introduce the results of our analysis. As a site vacancy, we remove one spin, and introduce a spin hole, as shown in Fig.~\ref{Fig.1}(a).
First, in Fig.~\ref{Fig.2} (f), we show the temperature dependence of specific heat, $C$. Reflecting the two energy scales in this system, $C$ takes double peak structure.
The high-$T$ ($T\sim0.4$) and low-$T$ ($T\sim0.01$) peaks correspond to the entropy release by c-Majorana and Vison, and they contribute to the entropy, $\frac{1}{2}\log2$ per spin, respectively~\cite{nasu2015thermal}. The plateau-like structure of entropy was, in fact, observed in RuCl$_3$~\cite{kubota2015successive} and in Na/Li$_2$IrO$_3$~\cite{mehlawat2017heat}.

In Fig.~\ref{Fig.2}(a), we show the local magnetic susceptibility, $\chi_i$, for $T=1.00, 0.10$ and $0.01$ along the two symmetry lines depicted in Fig.~\ref{Fig.1}(a) and Fig.~\ref{Fig.2}(e).
At $T=1.00$, $\chi_i$ is almost flat, without any spatial variation. As decreasing temperature, this tendency stays the same at $T=0.10$, while at $T=0.01$, only a slight enhancement appears around the vacancy.
This tendency is reflected in the temperature dependence shown in Fig.~\ref{Fig.2}(b).
Down to $T\sim0.02$, $\chi_i$ shows the same behavior as bulk value, which is obtained in the system without vacancy.
Only below $T\sim0.02$, several sites around the vacancy, show increasing behavior as lowering $T$,
possibly connecting to logarithmic growth predicted in the low $T$ region for fixed Z$_2$ flux sector~\cite{willans2010disorder,willans2011site}.

Next, let us look at $1/T_1$.
In Fig.~\ref{Fig.2}(c), we plot $\bigl(\frac{1}{T_1}\bigr)_j^{\alpha}$ along the symmetry lines as for $\chi_i$. 
In contrast to the static susceptibility, $1/T_1$ shows substantial spatial dependence from high temperature.
In particular, around $T=0.1$, the local $1/T_1$ exhibits
Friedel-like oscillation, and changes drastically around the vacancy site.
The oscillation sustains down to the lower temperature, $T=0.01$, while the magnitude of $1/T_1$ itself is suppressed.

We plot the $T$ dependence of $1/T_1$ for several sites [Fig.~\ref{Fig.2}(d)].
In contrast to $\chi_i$, the spatial differentiation of $1/T_1$ starts at rather high temperature.
Despite large spatial variation, the qualitative $T$ dependence do not differ between sites; they show a single peak around $T=0.02-0.1$.
However, an exception is found for the neighbor of vacancy. Specifically, the correlation of $S^{\alpha}$ in the $\alpha$-neighbor of the vacancy gives qualitatively different $T$ dependence of local $1/T_1$. As an example, we focus on $\bigl(\frac{1}{T_1}\bigr)_1^{x}$, at the site $1$ [Fig.~\ref{Fig.2} (e)], the $x$-neighbor of vacancy;
$\bigl(\frac{1}{T_1}\bigr)_1^{x}$ stays almost flat in a wide temperature range from high-$T$ limit down to $T\sim0.01$, then it shows a slight upturn around the temperature of low-$T$ Vison peak of specific heat [Fig.~\ref{Fig.2} (d)].
 
\begin{figure}[h]
\begin{center}
\includegraphics[width=0.51\textwidth]{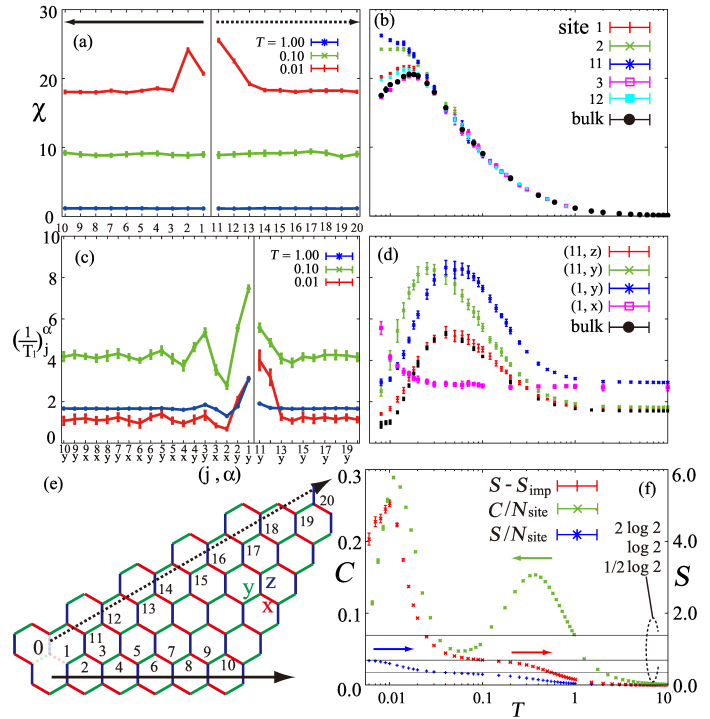}
\end{center}
\caption{\label{Fig.2} 
(color online). (a) (b) The local magnetic susceptibility, $\chi_{i}$, and (c) (d) $(1/T_1)_j^{\alpha}$. (e) Site number convention for (a)-(d). Site $0$ corresponds to the vacancy. (f) 
specific heat ($C$) and entropy ($S$) divided by the number of sites, $N_{\rm site}$. The entropy reduction by impurity ($\Delta S\equiv S-S_{\rm imp}$) is also shown here. Both $S$ and $S_{\rm imp}$ are extensive quantities, while $\Delta S$ is $\mathcal{O}(1)$.
(a) and (c) show the spatial variation of $\chi_{i}$ and $(1/T_1)_j^{\alpha}$ at $T=0.01$, $0.10$, and $1.00$, along the lines shown in (e) [See also Fig.~\ref{Fig.1}(a)]. The values along the solid (dashed) arrows in (e) are plotted in the lefthand (righthand) side.}
\end{figure}

\begin{figure}[h]
\begin{center}
\includegraphics[width=0.5\textwidth]{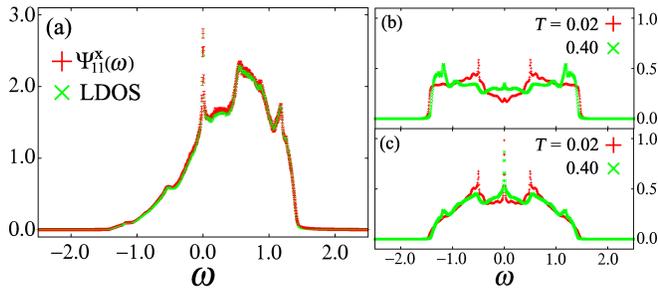}
\end{center}
\caption{\label{Fig.3} 
(color online). (a) The on-site spin correlation $\Psi_{11}^x(\omega)$ at the $x$-neighbor of the vacancy, site $1$, obtained at $T=0.4$. The red and green symbols are obtained from eq.~(\ref{analyticalformula}) and the scaled local density of states (\ref{scaled_LDOS}), respectively, and they give the same result. A sharp zero-energy peak appears at $\omega=0$.
(b) The density of states at $T=0.02$ and $0.40$ in a system without vacancy, (c) and at the $x$-neighbor of the vacancy.
}
\end{figure}

The anomalous temperature dependence of $1/T_1$ can be attributed to the zero-energy resonance in the spectrum of dynamical spin correlation.
As is clear in Fig.~\ref{Fig.3}(a), the local spectrum, $\Psi^{x}_{11}(\omega)$, at the neighborhood of the vacancy, shows a sharp zero-energy peak.

Resonance peak is elusive in a system with fractional excitations. 
In such systems, external perturbation usually gives rise to multi-particle process, which results in continuous rather than resonant spectra.
The zero-energy peak implies the two species of fractionalized excitations, Vison and c-Majorana, have zero mode simultaneously.
Indeed, we can explicitly show their existence.
To this aim, as a description of site vacancy, instead of removing a spin from site $0$, we assume the spin stays there, and set the couplings to neighboring spins to be zero.
From this viewpoint, starting from the ground state ($W_p=1$ everywhere), the operation of $S_1^x$ leads to a pair-creation of Visons on the both sides of the erased bond, connecting site $0$ and $1$ [Fig.~\ref{Fig.1}(c)]. However, this configuration can also be reached by operating $S_0^x$ at the ``removed" site, which, in fact, has no physical influence on the initial state.
This simple argument shows the pair-created Visons in Fig.~\ref{Fig.1}(c) is a zero mode.

Given that the pair-Vison does not have any physical consequences, the expression of $\Psi_{11}^x(\omega)$, given by eq.~(\ref{analyticalformula}), simplifies considerably.
By setting $A=A^{(j)}$ in eq.~(\ref{analyticalformula}), the correlation function reduces essentially to the local Majorana density of states:
\begin{eqnarray}
\hspace{-0.5cm}\Psi^{x}_{11} = \frac{2\pi}{1 + e^{-\beta\omega}}\frac{\sum_{\{W_p\}}Z[\{W_p\}]\sum_m\delta(\omega-\varepsilon^{(m)})|v^{(m)}_1|^2}{\sum_{\{W_p\}}Z[\{W_p\}]}
\label{scaled_LDOS}
\end{eqnarray}

We compare $\Psi^{x}_{11}(\omega)$ obtained from eq.~(\ref{analyticalformula}) and this scaled Majorana LDOS in Fig.~\ref{Fig.3}(a), and find a complete overlap as expected.
As comparison, we plot the bulk Majorana DOS in Fig.~\ref{Fig.3}(b).
We find no zero-energy peak in bulk DOS, and the weight at zero energy rather develops a dip as decreasing temperature, reflecting that the c-Majorana eventually has the same dispersion as Graphene at $T=0$, which has Dirac-like pseudo-gap.
In contrast, the zero-energy peak in LDOS next to vacancy sharpens, as decreasing temperature [Fig.~\ref{Fig.3}(c)].

The zero-energy peak in LDOS is attributed to the quasi-localized c-Majorana zero mode formed around the vacancy known for Graphene~\cite{pereira2006disorder}, and subsequently discussed in the context of Kitaev model in the fixed $Z_2$ flux sector~\cite{willans2010disorder,willans2011site}.
This c-Majorana zero mode interacts with thermally excited Visons at finite temperatures.
Visons act as scatters for c-Majoranas. So, one may expect they further localize the zero-energy state at high temperatures.
However, our numerical analysis implies contrary.
As lowering temperatures, the zero-energy peak sharpens [Fig.~\ref{Fig.3}(c)], i.e., the zero mode becomes more localized. 
The zero-energy state is quasi-localized by nature, so the Visons work in favor of delocalizing them.
In particular, below the low-$T$ peak of specific heat, the Vison becomes dilute. 
This population change leads to the higher zero-energy peak at lower temperatures, and consequently, the local $1/T_1$ shows upturn.
We also note the Friedel-like oscillation of $1/T_1$ can be attributed to this Majorana zero mode.
If Visons are absent, the zero mode has finite components on a different sublattice from the vacancy~\cite{pereira2006disorder}, where $1/T_1$ actually shows larger values [Fig.~\ref{Fig.2}(c)].

On experimental grounds, field-orientational dependence of $1/T_1$ also deserves attention.
the flat temperature dependence of local $1/T_1$ in high temperature region may be experimentally interpreted as a free spin localized around vacancy,
as is often the case with a system with broken spin singlet.
The distinction from this well-known case is found in the field-orientational dependence of local $1/T_1$.
So far, we have focused only on the $x$-neighbor of the vacancy, where the $x$-component of local correlation function gives anomalous $1/T_1$, which
can be observed only if magnetic field is applied perpendicular to $x$. 
Similarly, the zero-energy resonance appears for the $y$- and $z$-neighbors in $y$- and $z$-component of local correlation function.
The resultant field-orientational dependence of $1/T_1$ gives a clue to the identification of this anomalous zero-energy resonance.

Finally, we remark on the effect of site dilution on the specific heat. 
Without vacancy, the specific heat shows double peak structure, which gives rise to clear two-step plateaus in entropy [Fig.~\ref{Fig.2}(f)]; the high-$T$ (low-$T$) peak of specific heat results from the entropy release of $\frac{1}{2}\log 2$ per spin, by c-Majorana (Vison).
In Fig.~\ref{Fig.2}(f), we plot the entropy reduction by vacancy, $\Delta S\equiv S-S_{\rm imp}$, where $S_{\rm imp}$ and $S$ are the total entropy of the system with and without vacancy.  
$\Delta S$ shows the plateau at $\Delta S\sim\log 2$ at intermediate temperature.
This is a combination of two effect; Removal of one spin deprives $\frac{1}{2}\log 2$ of the high-$T$ c-Majorana peak.
Moreover, the zero mode formation requires one c-Majorana, which subtracts another $\frac{1}{2}\log 2$ from the high-$T$ peak.
Namely, this reduction of high-$T$ peak reflects the reconstruction of energy levels to recombine fractional excitations into a zero mode, which may be captured through a systematic study of site dilution.
Indeed, the high-$T$ peak of specific heat was observed in a couple of materials~\cite{kubota2015successive,mehlawat2017heat}.

To summarize, we have constructed a new analytical solution of dynamical correlation function of the Kitaev model.
By incorporating this solution into the classical Monte Carlo scheme, we fully analyzed the dynamical response of the system, especially, in the presence of site vacancy.
As a result, we found an enhancement of dynamical magnetic response in the vicinity of vacancy, which leads to Friedel-like oscillation in local $1/T_1$.
We further found a sharp zero-energy peak in the magnetic excitation spectrum, attributed to the Vison \& Majorana zero mode trapped near the site vacancy. 
This zero mode reflects the fractionalization of spin hole into an Ising triplet, and is observable through the characteristic temperature and field-orientational dependence of $1/T_1$.
The observation of zero-energy resonance will be useful to diagnose Kitaev's quantum spin liquid phase,
based on resonance rather than continuum, directly reflecting the nature of fractionalization.

We thank Tohru Koma for valuable discussions.
This work was supported by the JSPS KAKENHI (Nos. JP15H05852 and JP16H04026), MEXT, Japan. Part of numerical calculations were carried out
on the Supercomputer Center at Institute for Solid State Physics, University of Tokyo.


%

\newpage
\setcounter{figure}{0}
\setcounter{equation}{0}
\renewcommand{\bibnumfmt}[1]{[S#1]}
\renewcommand{\citenumfont}[1]{S#1}
\begin{widetext}
\begin{center}
\Large 
{\it Supplemental information}
\end{center}

\author{Masafumi Udagawa}%
\affiliation{%
Department of Physics, Gakushuin University, Mejiro, Toshima-ku, Tokyo 171-8588, Japan
}

\maketitle
\section{Derivation of f-parity formula}
\begin{figure}[h]
\begin{center}
\includegraphics[width=0.8\textwidth]{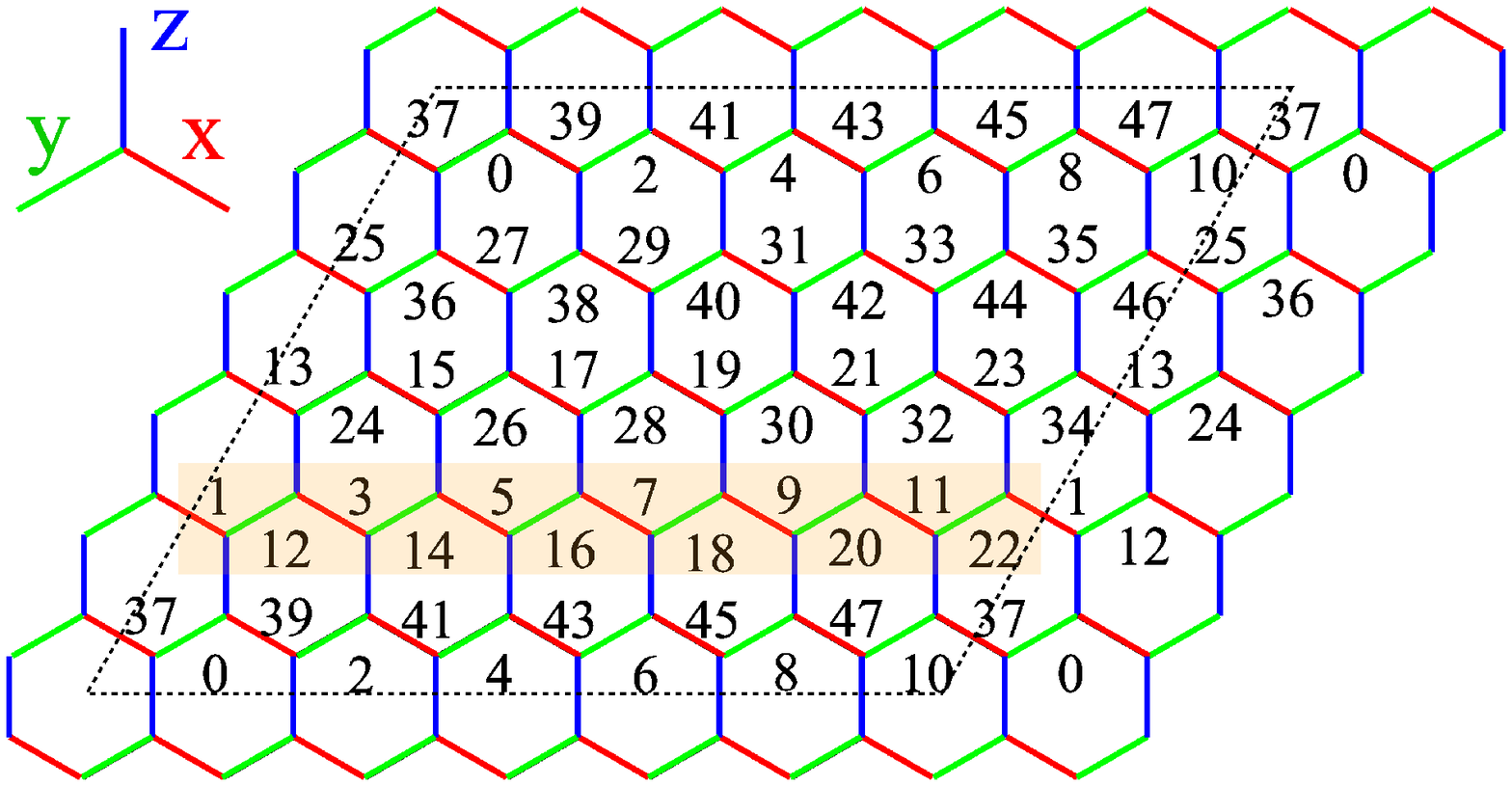}
\end{center}
\caption{\label{FIG.S1} 
(color online). Lattice geometry and numbering of sites for $N_1=6$ and $N_2=4$. 
Red, green and blue bonds host $s_i^{\alpha}s_{i_{\alpha}}^{\alpha}$-type couplings of $\alpha=x$, $y$ and $z$, respectively.
The second row is highlighted with orange shade for the explanation of the reordering of $b_j^{\alpha}$'s. See the text for details.
}
\end{figure}

Here, we introduce a simple derivation of f-parity formula, eq.~(3) in the main text, in a general setting.
We assume a system of $M\equiv N_1\times N_2$ unit cells under periodic boundary condition as depicted in FIG.~\ref{FIG.S1}.
The lattice geometry we introduced in the main text is a special case of $N_1=N_2=N$.
We adopt the convention for the numbering of sites as shown in FIG.~\ref{FIG.S1}. The sublattice $\Omega$ is composed of the even sites, $0, 2, \cdots 2M-2$. Accordingly, the product over the sublattice $\Omega$ can be denoted as $\prod_{i\in\Omega}u_i^{\alpha}=\prod_{m=0}^{M-1}u_{2m}^{\alpha}$.
The $x$-, $y$-, and $z$-neighbor of the site $2m$ is $2(m-N_1+1)+1$, $2(m-N_1)+1$ and $2m+1$ (modulo $2M$), respectively. Correspondingly, we define $u_{2m}^x=ib^x_{2m}b^x_{2(m-N_1+1)+1}$, $u_{2m}^y=ib^x_{2m}b^x_{2(m-N_1)+1}$, and $u_{2m}^z=ib_{2m}^{z}b_{2m+1}^{z}$.

As discussed by Pedrocci et al~\cite{Pedrocchi2011}, the physical fermion parity is determined from the projection onto the physical state:
\begin{eqnarray}
\prod_{j=0}^{2M-1}D_j\equiv\prod_{j=0}^{2M-1}b_j^xb_j^yb_j^zc_j = 1.
\end{eqnarray}
By reordering the operators, this relation is rewritten as
\begin{eqnarray}
(i^{M}\prod_{j=0}^{2M-1}b_j^x)(i^{M}\prod_{j=0}^{2M-1}b_j^y)(i^{M}\prod_{j=0}^{2M-1}b_j^z)(i^{M}\prod_{j=0}^{2M-1}c_j)=1.
\label{projection}
\end{eqnarray}
The product of $c_j$ can be transformed into fermion parity. To see this, note that the $z$-neighbor of site $2m$ corresponds to $2m+1$, in the numbering we adopted [FIG.~\ref{FIG.S1}]. Accordingly, following the definition of complex fermion in the main text, we set 
\begin{eqnarray}
c_{2m}=i(f_m^{\dag} - f_m), \ \ \ c_{2m+1}=f_m^{\dag} + f_m,
\label{complex_fermion}
\end{eqnarray}
which lead to
\begin{eqnarray}
i^{M}\prod_{j=0}^{2M-1}c_j = \prod_{m=0}^{M-1}(ic_{2m}c_{2m+1}) = \prod_{m=0}^{M-1}(1-2f_m^{\dag}f_m) = (-1)^F.
\end{eqnarray}
$(-1)^F$ counts the parity of the number of occupied fermions. Consequently, eq.~(\ref{projection}) is turned into
\begin{eqnarray}
(-1)^F = (i^{M}\prod_{j=0}^{2M-1}b_j^x)(i^{M}\prod_{j=0}^{2M-1}b_j^y)(i^{M}\prod_{j=0}^{2M-1}b_j^z).
\label{fparity_formula1}
\end{eqnarray}
Now, we rewrite the products of $b_j^{\alpha}$ into those of gauge fields. 
$\prod_{j=0}^{2M-1}b_j^{\alpha}=b_0^{\alpha}b_1^{\alpha}b_2^{\alpha}\cdots b_{2M-1}^{\alpha}$ is the product of $b_j^{\alpha}$ in the ordering of the numbers in FIG.~\ref{FIG.S1}. The reordering into the product of $u_{2m}^{\alpha}$ requires permutations of $b_j^{\alpha}$'s, which gives rise to sign factors. The product of $b_j^z$ is easiest to rewrite. Since the $z$-neighbor of site $2m$ is $2m+1$ as mentioned above, the successive pair in the product, $b_{2m}^zb_{2m+1}^z$ is simply transformed into $-iu_{2m}^z$. Then, we can rewrite
\begin{eqnarray}
i^{M}\prod_{j=0}^{2M-1}b_j^z = \prod_{m=0}^{M-1}u_{2m}^z.
\label{parity_z}
\end{eqnarray}
For the other two products, we rewrite 
\begin{eqnarray}
i^{M}\prod_{j=0}^{2M-1}b_j^x = (-1)^{\eta_x}\prod_{m=0}^{M-1}u_{2m}^x, \ \ \ i^{M}\prod_{j=0}^{2M-1}b_j^y = (-1)^{\eta_y}\prod_{m=0}^{M-1}u_{2m}^y,
\label{parity_xy}
\end{eqnarray}
where $\eta_x$ ($\eta_y$) means the parity of permutation that transforms the order of the product, $(b_0^{x(y)}, b_1^{x(y)}, \cdots b_{2M-1}^{x(y)})$ into the order of product of gauge field, $u_{2m}^x$ ($u_{2m}^y$). For example, in the special case of $N_1=6$ and $N_2=4$, the product of $u_{2m}^x$ looked like
\begin{eqnarray}
u_{12}^xu_{14}^xu_{16}^xu_{18}^xu_{20}^xu_{22}^x = i^6(b_{12}^xb_{3}^x)(b_{14}^xb_{5}^x)(b_{16}^xb_{7}^x)(b_{18}^xb_{9}^x)(b_{20}^xb_{11}^x)(b_{22}^xb_{1}^x),
\end{eqnarray}
in the second row of the FIG.~\ref{FIG.S1}, highlighted by the orange shade. Meanwhile,  the product of $u_{2m}^y$ in the same row looked like
\begin{eqnarray}
u_{12}^yu_{14}^yu_{16}^yu_{18}^yu_{20}^yu_{22}^y = i^6(b_{12}^yb_{1}^y)(b_{14}^yb_{3}^y)(b_{16}^yb_{5}^y)(b_{18}^yb_{7}^y)(b_{20}^yb_{9}^y)(b_{22}^yb_{11}^y).
\end{eqnarray}
The ordering of former is obtained from the latter by additional cyclic permutation: $(1, 3, 5, 7, 9, 11)\to(3, 5, 7, 9, 11, 1)$, which leads to the sign factor, $(-1)^{N_1-1}$.
Since each of $N_2$ rows gives the same factor, we found the relation:
\begin{eqnarray}
(-1)^{\eta_x} = (-1)^{\eta_y}(-1)^{N_2(N_1-1)}.
\label{parity_relation}
\end{eqnarray}
Combining eqs.~(\ref{fparity_formula1}), (\ref{parity_z}), (\ref{parity_xy}), (\ref{parity_relation}), we obtain
\begin{eqnarray}
(-1)^F =  (-1)^{N_2(N_1-1)}\prod_{m=0}^{M-1}u_{2m}^x\prod_{m=0}^{M-1}u_{2m}^y\prod_{m=0}^{M-1}u_{2m}^z.
\label{fparity_general}
\end{eqnarray}
As a special case: $N_1=N_2=N$, we obtain 
\begin{eqnarray}
(-1)^F =  \prod_{m=0}^{M-1}u_{2m}^x\prod_{m=0}^{M-1}u_{2m}^y\prod_{m=0}^{M-1}u_{2m}^z,
\label{fparity_special}
\end{eqnarray}
which leads to eq.~(3) in the main text.

\section{Taking fermion trace with f-parity projector}
In this section, we introduce how to take fermion trace of observables, in the presence of f-parity projector, $P_F=\frac{1}{2}(1 + (-1)^{\rm F_{ph}}e^{i\pi\sum_if_i^{\dag}f_i})$.

\subsection{Partition function}
Firstly, we start with partition function. Suppose we consider the Hamiltonian,
\begin{eqnarray}
\mathcal{H} = \frac{i}{4}c_kA_{kk'}c_{k'} 
= \frac{1}{2}\begin{bmatrix} f^{\dag} & f \end{bmatrix} K \begin{bmatrix} f \\ f^{\dag} \end{bmatrix} = \sum_{m=0}^{M-1}\varepsilon_m(\gamma_m^{\dag}\gamma_m-\frac{1}{2}),
\label{Majorana_diagonalization}
\end{eqnarray}
where $A$ is the $2M\times 2M$ anti-symmetric matrix. The eigenvalues of matrix $iA$ are composed of $M$ pairs of positive and negative energies, ($\varepsilon_m$, $-\varepsilon_m$) ($\varepsilon_m\geq0$, $m=0, 1, \cdots M-1$). $\hat{f} = (f_0, f_1, \cdots, f_{M-1})$ are a set of annihilation operators, defined by eq.~(\ref{complex_fermion}). $\gamma$'s are the fermion operators to diagonalize $K$, which are related to $\hat{f}$ by
\begin{eqnarray}
\hat{\gamma} = Q\hat{f}.
\end{eqnarray}
This equation relates the parity of $\gamma$-fermion, $(-1)^{F_{\gamma}}\equiv\prod_{m=0}^{M-1}(1-2\gamma_m^{\dag}\gamma_m)$, with that of f-fermions, as
\begin{eqnarray}
(-1)^{F_{\gamma}} = {\rm det}\ Q\ (-1)^F.
\end{eqnarray}
The expression (\ref{Majorana_diagonalization}) immediately leads to the partition function for a given $Z_2$ flux configurations, $\{W_p\}$,
\begin{eqnarray}
Z[\{W_p\}] = \prod_{m=0}^{M-1}(e^{\beta\frac{\varepsilon_m}{2}} + e^{-\beta\frac{\varepsilon_m}{2}}) = \prod_{m=0}^{M-1}\bigl(2\cosh\frac{\beta\varepsilon_m}{2}\bigr).
\label{partition_function}
\end{eqnarray}
Here, the occupied (unoccupied) $m$-th eigenstate contributes to the factor $e^{-\beta\frac{\varepsilon_m}{2}}$ ($e^{\beta\frac{\varepsilon_m}{2}}$). This observation immediately leads to the expression of f-parity weighted partition function,
\begin{align}
Z_f[\{W_p\}] &\equiv {\rm Tr}[e^{i\pi\sum_if_i^{\dag}f_i}e^{-\beta\mathcal{H}}] = {\rm det}\ Q{\rm Tr}[e^{i\pi\sum_m\gamma_m^{\dag}\gamma_m}e^{-\beta\mathcal{H}}]\nonumber\\
 &= {\rm det}\ Q\prod_{m=0}^{M-1}(e^{\beta\frac{\varepsilon_m}{2}} - e^{-\beta\frac{\varepsilon_m}{2}}) = {\rm det}\ Q\prod_{m=0}^{M-1}\bigl(2\sinh\frac{\beta\varepsilon_m}{2}\bigr).
\label{f_partition_function}
\end{align}
Combining eqs.~(\ref{partition_function}) and (\ref{f_partition_function}), and averaging over $\{W_p\}$, we obtain the partition function
\begin{align}
Z &= {\rm Tr}_{\{W_p\}}[P_Fe^{-\beta\mathcal{H}[\{W_p\}]}] = \frac{1}{2}\sum_{\{W_p\}}\Bigl(Z[\{W_p\}] + (-1)^{\rm F_{ph}}Z_f[\{W_p\}]\Bigr)\nonumber\\
 &= \frac{1}{2}\sum_{\{W_p\}}\bigl[\prod_{m=0}^{M-1}\bigl(2\cosh\frac{\beta\varepsilon_m}{2}\bigr) + (-1)^{\rm F_{ph}}{\rm det} Q\prod_{m=0}^{M-1}\bigl(2\sinh\frac{\beta\varepsilon_m}{2}\bigr)\bigr].
\label{full_partition_function}
\end{align}
From this, we can obtain physical observables, such as energy, $E$, and specific heat, $C$:
\begin{align}
E = \langle\mathcal{H}\rangle = -\frac{\sum_{\{W_p\}}Z[\{W_p\}](\sum_{m=0}^{M-1}\frac{\varepsilon_m}{2}\tanh(\frac{\beta\varepsilon_m}{2})) + (-1)^{\rm F_{ph}}Z_{f}[\{W_p\}](\sum_{m=0}^{M-1}\frac{\varepsilon_m}{2}\coth(\frac{\beta\varepsilon_m}{2}))}{\sum_{\{W_p\}}Z[\{W_p\}] + (-1)^{\rm F_{ph}}Z_{f}[\{W_p\}]},
\end{align}
and
\begin{align}
&C  = \frac{\langle\mathcal{H}^2\rangle - \langle\mathcal{H}\rangle^2}{T^2},\nonumber\\
&\langle\mathcal{H}^2\rangle = \frac{\sum_{\{W_p\}}Z[\{W_p\}](\sum_{m=0}^{M-1}(\frac{\varepsilon_m}{2})^2\frac{1}{\cosh^2(\frac{\beta\varepsilon_m}{2})}) - (-1)^{\rm F_{ph}}Z_{f}[\{W_p\}](\sum_{m=0}^{M-1}(\frac{\varepsilon_m}{2})^2\frac{1}{\sinh^2(\frac{\beta\varepsilon_m}{2})})}{\sum_{\{W_p\}}Z[\{W_p\}] +(-1)^{\rm F_{ph}}Z_{f}[\{W_p\}]}\nonumber\\
&+\frac{\sum_{\{W_p\}}Z[\{W_p\}](\sum_{m=0}^{M-1}\frac{\varepsilon_m}{2}\tanh(\frac{\beta\varepsilon_m}{2}))^2 + (-1)^{\rm F_{ph}}Z_{f}[\{W_p\}](\sum_{m=0}^{M-1}\frac{\varepsilon_m}{2}\coth(\frac{\beta\varepsilon_m}{2}))^2}{\sum_{\{W_p\}}Z[\{W_p\}] + (-1)^{\rm F_{ph}}Z^{\gamma}_{f}[\{W_p\}]}.
\end{align}

\section{The relation between $\Psi^{\alpha}_{jj'}(\omega)$ and observables}
In this section, we relate the correlation function we obtained,
\begin{eqnarray}
\delta_{\alpha\alpha'}\Psi_{jj'}^{\alpha}(\omega+i\delta) = \int_0^{\infty}dt\ e^{i(\omega + i\delta)t}\langle S_j^{\alpha}(t)S_{j'}^{\alpha'}(0)\rangle.
\end{eqnarray}
to the generalized magnetic susceptibility,
\begin{eqnarray}
\chi_{jj'}^{\alpha\alpha'}(\omega+i\delta) = i\int_0^{\infty}dt\ e^{i(\omega + i\delta)t}\langle[S_j^{\alpha}(t), S_{j'}^{\alpha'}(0)]\rangle,
\end{eqnarray}
which is directly associated with physical observables. By taking Lehmann representation, 
\begin{eqnarray}
\int_0^{\infty}dt\ e^{i(\omega + i\delta)t}\langle S_j^{\alpha}(t)S_{j'}^{\alpha'}(0)\rangle = \frac{i}{Z}\sum_{n,m}e^{-\beta E_n}\frac{\langle n|S_j^{\alpha}|m\rangle\langle m|S_{j'}^{\alpha'}|n\rangle}{\omega - (E_m-E_n) + i\delta}
\end{eqnarray}
For diagonal components: $\alpha=\alpha'$, by using time-reversal symmetry relation, we have
\begin{eqnarray}
\hspace{-0.5cm}\Psi_{jj'}^{\alpha}(\omega+i\delta) = \frac{i}{2Z}\sum_{n,m}\Bigl(\frac{e^{-\beta E_n}}{\omega - (E_m-E_n) + i\delta} + \frac{e^{-\beta E_m}}{\omega - (E_n-E_m) + i\delta}\Bigr)\langle n|S_j^{\alpha}|m\rangle\langle m|S_{j'}^{\alpha}|n\rangle.
\label{lehmann_psi}
\end{eqnarray}
Similarly, we have
\begin{eqnarray}
\chi_{jj'}^{\alpha\alpha'}(\omega+i\delta) = \frac{1}{Z}\sum_{n,m}(e^{-\beta E_m}-e^{-\beta E_n})\frac{\langle n|S_j^{\alpha}|m\rangle\langle m|S_{j'}^{\alpha'}|n\rangle}{\omega - (E_m-E_n) + i\delta}.
\label{lehmann_chi}
\end{eqnarray}
Comparing Eqs.~(\ref{lehmann_psi}) and (\ref{lehmann_chi}), we obtain the fluctuation-dissipation relation:
\begin{eqnarray}
{\rm Im}\ \chi_{jj'}^{\alpha\alpha}(\omega + i\delta) = 2\pi(1-e^{-\beta\omega}){\rm Re}\ \Psi_{jj'}^{\alpha}(\omega+i\delta),
\label{FDtheorem}
\end{eqnarray}
and for the real part, the $\omega=0$ components can be related via
\begin{eqnarray}
{\rm Re}\ \chi_{jj'}^{\alpha\alpha}(0) = -2{\rm Im}\ \Psi_{jj'}^{\alpha}(0+i\delta).
\label{dc_relation}
\end{eqnarray}

\subsection{Static susceptibility}
Local magnetic susceptibility, $\chi_i$ is defined as the local magnetization at site $i$, divided by the applied uniform magnetic field.
With magnetic field, ${\mathbf h} = (h_x, h_y, h_z)$, its diagonal component can be obtained as 
\begin{eqnarray}
\chi_i = (g\mu_{\rm B})^2\sum_{\alpha,\alpha'}\frac{h_{\alpha}h_{\alpha'}}{|{\mathbf h}|^2}\sum_{j}{\rm Re}\ \chi_{ij}^{\alpha\alpha'}(0)
\end{eqnarray}
according to Kubo formula. Consequently, for ${\mathbf h}\parallel[111]$, we have 
\begin{eqnarray}
\chi_i = -\frac{2}{3}(g\mu_{\rm B})^2\sum_{\alpha}{\rm Im}\ (\Psi_{ii}^{\alpha}(0+i\delta) + \Psi_{ii_{\alpha}}^{\alpha}(0+i\delta)).
\label{local_chi}
\end{eqnarray}
where we have used Eq.~(\ref{dc_relation}), and the relation that $\Psi_{ij}^{\alpha}(\omega+i\delta)=\delta_{ji}\Psi_{ii}^{\alpha}(\omega+i\delta) + \delta_{ji_{\alpha}}\Psi_{ji_{\alpha}}^{\alpha}(\omega+i\delta)$.

\subsection{Spin-lattice relaxation rate}
The local magnetic susceptibility plays a key role in the spin-lattice relaxation rate, $1/T_1$. $1/T_1$ of a certain nucleus spin is obtained from the correlation function of the local fields, which is expressed by
\begin{eqnarray}
\frac{1}{T_1} = T\sum_{j,j'}A^{\alpha,\alpha'}_{j,j'}\lim_{\omega\to0}\frac{{\rm Im}\chi^{(\perp)\alpha\alpha'}_{jj'}(\omega)}{\omega},
\end{eqnarray}
taking account of the contributions from neighboring sites. In this work, we use 
\begin{eqnarray}
\Bigl(\frac{1}{T_1}\Bigr)_j^{\alpha}\equiv \frac{T}{2\pi}\lim_{\omega\to0}\frac{{\rm Im}\chi^{\alpha\alpha}_{jj}(\omega)}{\omega},
\end{eqnarray}
as an estimate of $1/T_1$ in the vicinity of site $j$, assuming the magnetic field is applied perpendicular to $\alpha$. From Eq.~(\ref{FDtheorem}), one can relate this to the correlation function, $\Psi_{jj'}^{\alpha}(\omega+i\delta)$ as
\begin{eqnarray}
\Bigl(\frac{1}{T_1}\Bigr)_j^{\alpha} = {\rm Re}\ \Psi_{jj}^{\alpha}(0+i\delta).
\label{local_T1}
\end{eqnarray}

\section{Derivation of the analytical solution of $\langle S_j^{\alpha}(t)S_{j'}^{\alpha}(0)\rangle$}
Here, we give the derivation of the analytical solution of $\langle S_j^{\alpha}(t)S_{j'}^{\alpha}(0)\rangle$. 

\subsection{Useful operator relations}
Before introducing derivation, here we give several useful operator relations. For skew-symmetric (but not necessarily Hermitian) matrix, $iA$ and $iB$, and a set of Majorana operators, $\{c_j\}$, the following relations hold.\\
$[$1$]$ Product:
\begin{align}
e^{\frac{i}{4}c_kA_{kk'}c_{k'}}e^{\frac{i}{4}c_kB_{kk'}c_{k'}} = e^{\frac{i}{4}c_kD_{kk'}c_{k'}}, 
\end{align}
where $D$ is a skew-symmetric matrix which satisfies $e^{iD}=e^{iA}e^{iB}$.

\noindent
$[$2$]$ Trace:
\begin{align}
{\rm Tr}[e^{\frac{i}{4}c_kA_{kk'}c_{k'}}] = \sqrt{{\rm det}(1 + e^{iA})}.
\end{align}
As a corollary, combining with $[$1$]$, 
\begin{align}
{\rm Tr}[e^{\frac{i}{4}c_kA_{kk'}c_{k'}}e^{\frac{i}{4}c_kB_{kk'}c_{k'}}] = \sqrt{{\rm det}(1 + e^{iA}e^{iB})}.
\end{align}

\noindent
$[$3$]$ Correlator:
\begin{align}
\frac{{\rm Tr}[e^{\frac{i}{4}c_kA_{kk'}c_{k'}}c_jc_{j'}]}{{\rm Tr}[e^{\frac{i}{4}c_kA_{kk'}c_{k'}}]} = \Bigl[\frac{2}{1+e^{iA}}\Bigr]_{jj'}.
\end{align}
As a corollary, combining with $[$1$]$, 
\begin{align}
\frac{{\rm Tr}[e^{\frac{i}{4}c_kA_{kk'}c_{k'}}e^{\frac{i}{4}c_kB_{kk'}c_{k'}}c_jc_{j'}]}{{\rm Tr}[e^{\frac{i}{4}c_kA_{kk'}c_{k'}}e^{\frac{i}{4}c_kB_{kk'}c_{k'}}]} = \Bigl[\frac{2}{1+e^{iA}e^{iB}}\Bigr]_{jj'}.
\end{align}

\noindent
$[$4$]$ f-parity operator:
\begin{align}
e^{i\pi\sum_{j=0}^{N-1}f_j^{\dag}f_j} = \prod_{j=0}^{N-1}e^{i\frac{\pi}{2}(1-ic_{2j}c_{2j+1})} = i^Ne^{\frac{i}{4}c_k\Sigma^y_{kk'}c_{k'}},
\end{align}
where $\Sigma^y = \pi\otimes_{j=0}^{N-1}\sigma^{y(2j,2j+1)}$, with $\sigma^{y(2j,2j+1)}$, the block Pauli matrix spanned by $2j$ and $2j+1$ columns and rows.
$e^{i\Sigma^y}=-1$ follows from this expression.

\subsection{Derivation}
Without loss of generality, we consider the case of $\alpha=z$, and we assume $j\in\Omega$.
Firstly, by adopting the Majorana expression of the spin operator: $S_j^{\alpha} = \frac{i}{2}b_j^{\alpha}c_j$, we obtain
\begin{align}
&\langle S_j^{z}(t)S_{j'}^{z}(0)\rangle = -\frac{1}{4}\frac{\sum_{\{W_p\}}{\rm Tr}[P_F e^{-\beta\frac{i}{4}c_kA_{kk'}c_{k'}}b_j^z(t)c_j(t)b_{j'}^z(0)c_{j'}(0)]}{\sum_{\{W_p\}}{\rm Tr}[P_F e^{-\beta\frac{i}{4}c_kA_{kk'}c_{k'}}]}\nonumber\\
&=-\frac{1}{4}\frac{\sum_{\{W_p\}}Z[\{W_p\}]\frac{{\rm Tr}[e^{-\beta\frac{i}{4}c_kA_{kk'}c_{k'}}b_j^z(t)c_j(t)b_{j'}^zc_{j'}]}{Z[\{W_p\}]} + (-1)^{\rm F_{ph}}Z_f[\{W_p\}]\frac{{\rm Tr}[e^{i\pi\sum_jf_j^{\dag}f_j}e^{-\beta\frac{i}{4}c_kA_{kk'}c_{k'}}b_j^z(t)c_j(t)b_{j'}^zc_{j'}]}{Z_f[\{W_p\}]}}{\sum_{\{W_p\}}Z[\{W_p\}] + (-1)^FZ_f[\{W_p\}]},
\label{correlation_equation}
\end{align}
where $(-1)^F$ is the physical f-parity given by eq.~(\ref{fparity_general}) or (\ref{fparity_special}).
$Z[\{W_p\}]$ and $Z_f[\{W_p\}]$ are normal and f-parity weighted partition function defined by (\ref{partition_function}) and (\ref{f_partition_function}), respectively.
These partition functions are also written as
\begin{align}
Z[\{W_p\}] = \sqrt{{\rm det}(1+e^{-\beta iA})}, \ \ \ Z_f[\{W_p\}] = \sqrt{{\rm det}(1-e^{-\beta iA})}.
\end{align}
Here, we focus on the first part of the numerator:
\begin{align}
&\frac{{\rm Tr}[e^{-\beta\frac{i}{4}c_kA_{kk'}c_{k'}}b_j^z(t)c_j(t)b_{j'}^zc_{j'}]}{Z[\{W_p\}]} = \frac{{\rm Tr}[e^{-(\beta-it)\frac{i}{4}c_kA_{kk'}c_{k'}}b_j^zc_je^{-it\frac{i}{4}c_kA_{kk'}c_{k'}}b_{j'}^zc_{j'}]}{Z[\{W_p\}]}\nonumber\\
&=-\frac{{\rm Tr}[c_{j''}e^{-(\beta-it)\frac{i}{4}c_kA_{kk'}c_{k'}}b_j^ze^{-it\frac{i}{4}c_kA_{kk'}c_{k'}}b_{j'}^zc_{j'}]}{Z[\{W_p\}]}[e^{-(\beta-it)iA}]_{j''j}\nonumber\\
&=-\frac{{\rm Tr}[e^{-(\beta-it)\frac{i}{4}c_kA_{kk'}c_{k'}}e^{-it\frac{i}{4}c_kA^{(j)}_{kk'}c_{k'}}b_j^zb_{j'}^zc_{j'}c_{j''}]}{Z[\{W_p\}]}[e^{-(\beta-it)iA}]_{j''j},
\label{derivation_firstpart}
\end{align}
Here we have used
\begin{eqnarray}
b_j^ze^{-it\frac{i}{4}c_kA_{kk'}c_{k'}} = e^{-it\frac{i}{4}c_kA^{(j)}_{kk'}c_{k'}}b_j^z,
\end{eqnarray}
where $A^{(j)}$ is obtained from $A$ by reversing the sign of gauge field, $u^z_j$.
\begin{align}
(\ref{derivation_firstpart})&=-\frac{{\rm Tr}[e^{-(\beta-it)\frac{i}{4}c_kA_{kk'}c_{k'}}e^{-it\frac{i}{4}c_kA^{(j)}_{kk'}c_{k'}}c_{j'}c_{j''}]}{Z[\{W_p\}]}[e^{-(\beta-it)iA}]_{j''j}(\delta_{jj'} - iu_j^z\delta_{j'j_{z}})\nonumber\\
&=-\frac{{\rm Tr}[e^{-(\beta-it)\frac{i}{4}c_kA_{kk'}c_{k'}}e^{-it\frac{i}{4}c_kA^{(j)}_{kk'}c_{k'}}]}{Z[\{W_p\}]}
\frac{{\rm Tr}[e^{-(\beta-it)\frac{i}{4}c_kA_{kk'}c_{k'}}e^{-it\frac{i}{4}c_kA^{(j)}_{kk'}c_{k'}}c_{j'}c_{j''}]}{{\rm Tr}[e^{-(\beta-it)\frac{i}{4}c_kA_{kk'}c_{k'}}e^{-it\frac{i}{4}c_kA^{(j)}_{kk'}c_{k'}}]}[e^{-(\beta-it)iA}]_{j''j}(\delta_{jj'} - iu_j^z\delta_{j'j_{z}})\nonumber\\
&=-\sqrt{{\rm det}\Bigl(\frac{1+e^{-(\beta-it)iA}e^{-it\cdot iA^{(j)}}}{1+e^{-\beta iA}}\Bigr)}\Bigl[\frac{2}{1+e^{-(\beta-it)iA}e^{-it\cdot iA^{(j)}}}\Bigr]_{j'j''}[e^{-(\beta-it)iA}]_{j''j}(\delta_{jj'} - iu_j^z\delta_{j'j_{z}})\nonumber\\
&=-2\sqrt{{\rm det}\Bigl(\frac{1+e^{-(\beta-it)iA}e^{-it\cdot iA^{(j)}}}{1+e^{-\beta iA}}\Bigr)}\Bigl[\frac{1}{1+e^{-(\beta-it)iA}e^{-it\cdot iA^{(j)}}}e^{-(\beta-it)iA}\Bigr]_{j'j}(\delta_{jj'} - iu_j^z\delta_{j'j_{z}}).
\end{align}
Similarly, the second part can be simplified as
\begin{align}
&\frac{{\rm Tr}[e^{i\pi\sum_jf_j^{\dag}f_j}e^{-\beta\frac{i}{4}c_kA_{kk'}c_{k'}}b_j^z(t)c_j(t)b_{j'}^zc_{j'}]}{Z_f[\{W_p\}]}\nonumber\\
&=-\frac{{\rm Tr}[e^{i\pi\sum_jf_j^{\dag}f_j}e^{-(\beta-it)\frac{i}{4}c_kA_{kk'}c_{k'}}e^{-it\frac{i}{4}c_kA^{(j)}_{kk'}c_{k'}}c_{j'}c_{j''}]}{Z_f[\{W_p\}]}[e^{i\Sigma^y}e^{-(\beta-it)iA}]_{j''j}(\delta_{jj'} - iu_j^z\delta_{j'j_{z}})\nonumber\\
&=-2\sqrt{{\rm det}\Bigl(\frac{1+e^{i\Sigma^y}e^{-(\beta-it)iA}e^{-it\cdot iA^{(j)}}}{1-e^{-\beta iA}}\Bigr)}\Bigl[\frac{1}{1+e^{i\Sigma^y}e^{-(\beta-it)iA}e^{-it\cdot iA^{(j)}}}e^{i\Sigma^y}e^{-(\beta-it)iA}\Bigr]_{j'j}(\delta_{jj'} - iu_j^z\delta_{j'j_{z}})\nonumber\\
&=2\sqrt{{\rm det}\Bigl(\frac{1-e^{-(\beta-it)iA}e^{-it\cdot iA^{(j)}}}{1-e^{-\beta iA}}\Bigr)}\Bigl[\frac{1}{1-e^{-(\beta-it)iA}e^{-it\cdot iA^{(j)}}}e^{-(\beta-it)iA}\Bigr]_{j'j}(\delta_{jj'} - iu_j^z\delta_{j'j_{z}}).
\label{derivation_secondpart}
\end{align}
Putting eqs.~(\ref{derivation_firstpart}) and (\ref{derivation_secondpart}) into (\ref{correlation_equation}), we obtain the analytical solution eq.~(9) in the main text.

\section{Numerical procedure to obtain $\Psi^{\alpha}_{jj'}(\omega)$}
To obtain $\Psi^{\alpha}_{jj'}(\omega)$, we first calculate the time-dependent correlation function, $\Psi_{jj'}^{\alpha}(t)\equiv\langle S_j^{\alpha}(t)S_{j'}^{\alpha}(0)\rangle$, and make a numerical Fourier transformation to real-frequency representation. In FIG.~\ref{FIG.S2}, we show the results of the system without site vacancy for the system size, $N=12$. $\Psi_{jj'}^{\alpha}(t)$ shows a rapid oscillation in short time scale, then a slow decay takes over in a longer time scale. In FIG.~\ref{FIG.S2}, we show the on-site component of $\Psi_{jj'}^{\alpha}(t)$ for $T=0.01, 0.02, 0.10, 0.40$ and $1.00$. As temperature is lowered, the initial-time oscillation becomes clear and tends to sustain for a longer time. At the same time, the correlation stays finite for a longer time at lower temperatures: at $T=1.00$, the correlation becomes negligible before $t=10$, however, at $T=0.01$ it stays around $t=100$. On computational grounds, this means we need to calculate for a longer time at lower temperatures. At $T=0.01$, which is close to the lowest temperatures in this work, $t=200$ is enough to obtain convergent result. We use this value as the cutoff time as we mentioned in the main text.

\begin{figure}[h]
\begin{center}
\includegraphics[width=0.65\textwidth]{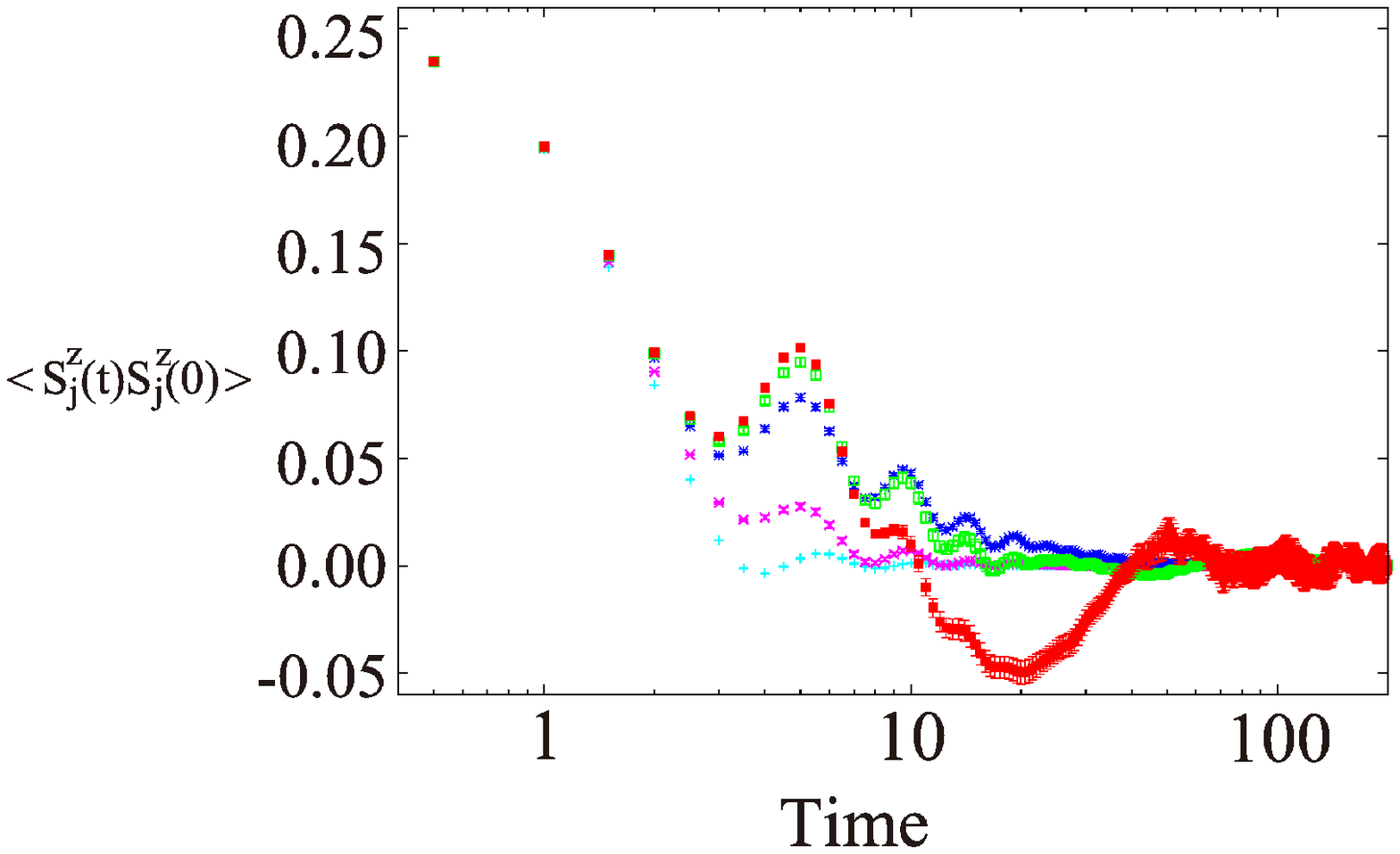}
\end{center}
\caption{\label{FIG.S2} 
(color online). The time dependence of the on-site correlation function, $\Psi_{jj'}^{\alpha}(t)\equiv\langle S_j^{\alpha}(t)S_{j'}^{\alpha}(0)\rangle$ for $T=0.01, 0.02, 0.10, 0.40$ and $1.00$ as shown in red, blue, green, purple, and light blue symbols.
}
\end{figure}

\begin{figure}[h]
\begin{center}
\includegraphics[width=0.8\textwidth]{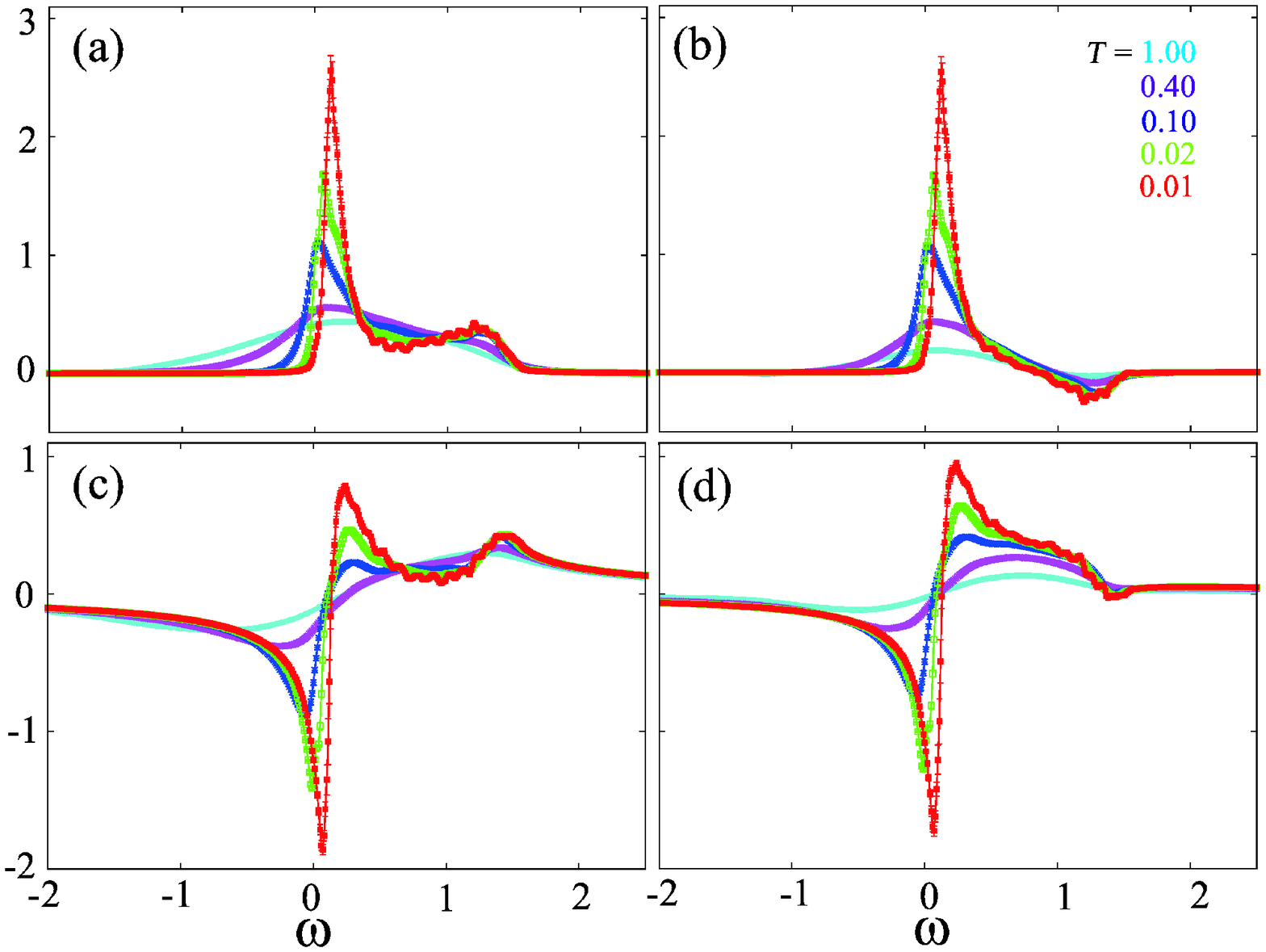}
\end{center}
\caption{\label{FIG.S2_2} 
(color online). The frequency dependence of the dynamical correlation functions. (a) (c) the on-site and (b) (d) off-site components of $\Psi_{jj'}^{\alpha}(\omega)$ are plotted for $T=0.01, 0.02, 0.10, 0.40$ and $1.00$. (a) and (b) show the real part, while (c) and (d) show the imaginary components, whose zero-frequency parts, as shown with vertical thin lines, are connected with spin-lattice relaxation rates, and magnetic susceptibilities, by eqs.~(\ref{local_T1}) and (\ref{local_chi}), respectively.
}
\end{figure}

In FIG.~\ref{FIG.S2_2}, we plott the frequency representation of correlation function, $\Psi_{jj'}^{\alpha}(\omega)$ for the system without site vacancy. We show the on-site ($j'=j$) in (a) and (c), while the off-site ($j'=j_{\alpha}$) components in (b) and (d), respectively. (a) and (b) show the real parts, and (c) and (d) show the imaginary parts. The real parts of the spectra show broad $\omega$ dependence at higher temperatures. As decreasing temperatures, we can observe two qualitative changes. Firstly, a broad continuum is formed at high energy part up to $\sim J$. This high-energy continuum shows opposite signs between the on-site and off-site components. Secondly, as decreasing temperature further, a sharp quasi-resonance peak is developed at low energy. This peak is not located exactly at zero energy, but at slightly finite energy of the order of flux gap. 

Correspondingly, the imaginary parts develop broad bump around the same energy as the high-energy continuum in the real parts, as well as
the sharp structures at low energies, as decreasing temperatures. These spectral features are well consistent the results of previous studies~\cite{Yoshitake2016}.
The zero-energy components of real and imaginary parts as shown with the vertical thin lines in FIG.~\ref{FIG.S2_2} are connected with the experimental observables, according to eqs.~(\ref{local_T1}) and (\ref{local_chi}), respectively.

\end{widetext}

\end{document}